%% file: rota.tex
\documentstyle[preprint,aps]{revtex}

\input climacro

\begin{document}

\title{\bf ON THE EQUATION $\nabla \times \mbox{\boldmath $a$}
= \kappa \mbox{\boldmath $a$} $}

\author{J. Vaz, Jr.\footnote{vaz@ime.unicamp.br}
and W. A. Rodrigues, Jr.\footnote{walrod@ime.unicamp.br}}

\address{Departamento de Matem\'atica Aplicada - IMECC - UNICAMP \\
CP 6065, 13081-970 Campinas, SP, Brazil}

\maketitle 

\begin{abstract}
\noindent We show that when correctly formulated the equation 
$\nabla \times \mbox{\boldmath $a$} = \kappa 
\mbox{\boldmath $a$}$ does not exhibit some 
inconsistencies atributed to it, so that its 
solutions can represent physical fields. 
\end{abstract}

\vspace{2ex} 


\vspace{3ex}

\noindent Let us consider the {\it free\/} Maxwell equations: 
\begin{eqnarray}
&{}& \nabla\cdot \vec{E} = 0 , \quad 
\nabla\cdot\vec{B} = 0 , \\
&{}& \nabla\times\vec{E} = -\frac{\partial\vec{B}}{\partial t} , 
\quad \nabla\times\vec{B} = \frac{\partial\vec{E}}{\partial t} .
\end{eqnarray}
We want to look for solutions of Maxwell equations which 
describe {\it stationary\/} electromagnetic configurations -- 
in the sense that the energy of the field does not 
propagate. 
In order to obtain one such stationary solution it is sufficient
to find solutions of the vector equation 
\begin{equation}
\label{eq.2}
\nabla \times \vec{a} = \kappa \vec{a} , \quad \kappa \quad {\rm constant} .
\end{equation}
In fact, if we are looking for stationary solutions then in the rest 
frame we can make the following {\it ansatz}: 
\begin{equation}
\label{eq.3}
\vec{E} = \vec{a}\sin{\kappa t} , 
\quad \vec{B} = \vec{a}\cos{\kappa t} .
\end{equation}
All Maxwell equations are automatically satisfied 
within this {\it ansatz\/} for $\vec{a}$ satisfying the 
vector equation (\ref{eq.2}). The solution is obviously 
stationary since the Poynting vector $\vec{S} = \vec{E} 
\times \vec{B} = 0$. It also follows that 
$\vec{E}$ and $\vec{B}$ satisfy the same equation:
\begin{equation}
\label{eq.4}
\nabla\times\vec{E} =\kappa\vec{E} , \quad 
\nabla\times\vec{B} = \kappa\vec{B} .
\end{equation}
The vector equation $\nabla\times\vec{B} = \kappa\vec{B}$ 
is very important in plasma physics and astrophysics, and 
can also be used as a model for force-free electromagnetic 
waves \cite{examples}. 

The identification of solutions of the eq.(\ref{eq.2}) 
with physical fields (as in eq.(\ref{eq.3}) above) 
has been criticized by Salingaros \cite{Saling}. 
In particular, he discussed the question of violation 
of gauge invariance and of parity invariance. The inconsistencies 
have been identified in \cite{Saling} with the 
lack of covariance of the eq.(\ref{eq.2}) with 
respect to transformations. Our proposal in this 
letter is to show that there is {\it no\/} violation of 
gauge invariance and of parity invariance. 

The argument leading to the lack of gauge invariance \cite{Saling}
runs as follows. From $\nabla\times\vec{B} = \kappa\vec{B}$ 
we have, since $\vec{B} = \nabla\times\vec{A}$, that 
$\nabla\times\vec{B} = \kappa\nabla\times\vec{A}$, and 
then $\vec{B} = \kappa\vec{A} + \nabla\phi$. Now, in 
\cite{Saling} it was argued that for $\vec{B}^\prime = 
\kappa \vec{A}^\prime + \nabla\phi = 
\kappa(\vec{A} + \nabla\lambda) + \nabla\phi = 
\vec{B} + \kappa\nabla\lambda$, that is, gauge 
invariance requires $\kappa = 0$ or the specific gauge 
$\lambda = 0$. The mistake in this argument is easily 
identified since for $\vec{B}^\prime = \nabla\times 
\vec{A}^\prime$ we have $\vec{B} = \kappa\vec{A}^\prime 
+ \nabla\psi$, where the arbitrary function $\psi$ 
must {\it not\/} be identified {\it a priori\/} 
with $\phi$. In this case $\vec{B}^\prime = 
\kappa(\vec{A}+\nabla\lambda)+\nabla\psi = 
\kappa\vec{A} + \kappa\nabla(\lambda+\psi) = 
\kappa\vec{A} + \kappa\nabla\phi = \vec{B}$. 

The argument used in \cite{Saling} leading to the lack of parity 
invariance is that since $\vec{B}$ is a parity eigenvector of
even parity \cite{Landau} and since under 
upon reflection we have $\nabla \mapsto -\nabla$ then 
$\kappa\vec{B} = \nabla\times\vec{B} \mapsto 
\kappa\vec{B} = -\nabla\times\vec{B}$, 
$\vec{B} = -\vec{B} = 0$, which means that solutions 
of $\nabla\times\vec{B} = \kappa\vec{B}$ must necessarily {\it not\/}
be a parity eigenvector, and then they cannot be associated 
with neither $\vec{E}$ nor $\vec{B}$ since both fields have 
definite parity. The origin of the mistake in this case 
is not trivial, and requires a detailed explanation. 

The problem in the above argument is essentially due to the 
definition of the vector product $\times$ in the usual 
Gibbs-Heaviside vector algebra. The usual definition of 
the vector product $\vec{v}\times\vec{u}$ as 
\begin{equation}
\label{eq.5}
(v_1,v_2,v_3)\times(u_1,u_2,u_3) = 
(v_2u_3-v_3u_2,v_3u_1-v_1u_3,v_1u_2-v_2u_1) 
\end{equation}
is a {\it nonsense\/} since it equals a pseudo-vector (L.H.S.) 
and a vector (R.H.S.). This nonsense is therefore also expected 
in the definition of $\nabla\times\vec{v}$: 
\begin{equation}
\label{eq.6}
\nabla\times\vec{v}=\left(\frac{\partial v_3}{\partial x_2} - 
\frac{\partial v_2}{\partial x_3}, 
\frac{\partial v_1}{\partial x_3} - \frac{\partial v_3}{\partial x_1}, 
\frac{\partial v_2}{\partial x_1} - \frac{\partial v_2}{\partial x_2}
\right) .
\end{equation}
In other words, in the Gibbs-Heaviside vector algebra the 
vector product of two vector $\vec{v}, \vec{u} \in V \simeq \BR^3$ 
is the mapping $\times : (\vec{v},\vec{u}) \mapsto \vec{w}$. 
Obviously $\vec{w}$ cannot belong to the same space $V$ 
where $\vec{v}$ and $\vec{u}$ live because $\vec{w}$ is a 
pseudo-vector. So, let us call this new vector space 
$V^\times$. We also have the vector product of vectors 
and pseudo-vectors, $\times : V \times V^\times \rightarrow V$ 
and $\times : V^\times \times V \rightarrow V$. The non-specification 
of these two spaces $V$ and $V^\times$ in the usual 
presentation produces nonsense. If we usually identify 
$V$ and $V^\times$ as in eq.(\ref{eq.5}) and consider the 
sum $\vec{v} + \vec{v}^\times = \vec{z}$, then under reflection 
is $\vec{z}$ a vector or a pseudo-vector? Obviously this means 
that the usual vector product is a nonsense. 

One formalism we can use which is free from  the above inconsistency 
is the one of {\it differential forms\/} \cite{Schultz}, or 
the Cartan calculus. Given the 1-forms $\{d x^i\}$ ($i = 1,2,3$) 
and the vector fields $\{\partial_j = \partial /\partial x^j\}$ 
($j = 1,2,3$) such that 
\begin{equation}
\label{eq.7}
\partial_j \JJ dx^i = 
dx^i (\partial_j) = \delta^i_j , 
\end{equation}
we can construct 1-forms $\bv$ and $\bu$ as
\begin{equation}
\label{eq.8}
\bv = v_i dx^i , \quad \bu = u_i dx^i .
\end{equation}
The exterior product gives the 2-form 
\begin{equation}
\label{eq.9}
\bv \wedge \bu = 
(v_1u_2-v_2u_1)dx^1\wedge dx^2 + 
(v_2u_3-v_3u_2)dx^2\wedge dx^3 + 
(v_1u_3 - v_3u_1)dx^1 \wedge dx^3 .
\end{equation}
In order to relate this expression with the vector product 
we need the so called Hodge operator $\star$ \cite{Schultz}. 
If we denote the volume element by $\tau$, 
\begin{equation}
\label{eq.10}
\tau = dx^1 \wedge dx^2 \wedge dx^3 
\end{equation}
then we have that 
\begin{equation}
\label{eq.11}
\star (\bv\w\bu\w\cdots\w\bw) = 
\vec{w}\JJ(\cdots\JJ(\vec{u}\JJ(\vec{v}\JJ\tau))\cdots) , 
\end{equation}
where $\vec{v} = \varphi(\bv)$, etc., and $\varphi$ is 
the isomorphism  given by 
\begin{equation}
\label{eq.12}
\varphi(dx^i) = \partial_i .
\end{equation}
Explicitly we have 
\begin{eqnarray}
\label{eq.13}
&{}& \star dx^1 = dx^2\w dx^3 , \quad 
\star dx^2 = dx^3 \w dx^1 , \quad 
\star dx^3 = dx^1 \w dx^2 , \\
\label{eq.14}
&{}& \star (dx^2 \w dx^3) = dx^1 , \quad 
\star (dx^3 \w dx^1) = dx^2 , \quad 
\star (dx^1 \w dx^2) = dx^3 .
\end{eqnarray}
It follows that $\star(\bv\w\bu)$ is the 1-form 
\begin{equation}
\label{eq.15}
\star(\bv\w\bu) = 
(v_2u_3-v_3u_2)dx^1 + 
(v_3u_1-v_1u_3)dx^2 + 
(v_1u_2-v_2u_1)dx^3 , 
\end{equation}
which we recognize as the counterpart of the vector product. 
If we work with $\star(\bv\w\bu)$ then if we take 
$dx^i \mapsto -dx^i$ we have $\star(\bv\w\bu) 
\mapsto -\star(\bv\w\bu)$ while $\bv\w\bu \mapsto \bv\w\bu$. 
This is because the volume element $\tau$ used in the 
definition of $\star$ also changes sign, $\tau \mapsto -\tau$. 

Now, the electric field is represented by a 1-form $\bE$ 
given by 
\begin{equation}
\label{eq.16}
\bE = E_1dx^1+E_2dx^2 + E_3dx^3 , 
\end{equation}
but the magnetic field is represented by a 
2-form $\bB$ 
\begin{equation}
\label{eq.17}
\bB = B_1 dx^2\w dx^3 + B_2 dx^3\w dx^1 + B_3 dx^1 \w dx^2 .
\end{equation}
The fact that the magnetic field {\it must\/} be represented 
by a 2-form follows from Faraday law of 
induction \cite{Westenholz}. Note that for $dx^i \mapsto 
-dx^i$ we have $\bE \mapsto -\bE$ and $\bB \mapsto \bB$. 
Note also that we can define a 1-form $\bb$ by 
\begin{equation}
\label{eq.18}
\bb = \star \bB = B_1dx^1+B_2dx^2+B_3dx^3 , 
\end{equation}
and in this case $\bb \mapsto -\bb$ for $dx^i \mapsto -dx^i$. 

Consider the differential operator $d$, which can be defined 
by 
\begin{equation}
\label{eq.19}
d\bv = \partial_i v_j dx^i\w dx^j , \quad 
d(\bv\w\bu) = (d\bv)\w\bu - \bv\w(d\bu) . 
\end{equation}
The codifferential operator $\delta$ is defined as 
\begin{equation}
\label{eq.20}
\delta = \star d \star .
\end{equation}
We can easily verify the relations 
\begin{eqnarray}
&{}& \nabla\times\vec{E} \leftrightarrow 
\star d \bE , \nonumber \\
&{}& \nabla \cdot \vec{E} \leftrightarrow 
\delta \bE , \nonumber \\
&{}& \nabla \times \vec{B} \leftrightarrow 
\delta \bB , \nonumber \\
\label{eq.21}
&{}& \nabla \cdot \vec{B} \leftrightarrow 
\star d \bB .
\end{eqnarray}
The vector equation $\nabla\times\vec{B} = \kappa\vec{B}$ must 
be written as 
\begin{equation}
\label{eq.22}
\delta \bB = \kappa \star \bB .
\end{equation}
The operators $d$ and $\delta$ are such that $d \mapsto -d$ and 
$\delta \mapsto -\delta$ for $dx^i \mapsto -dx^i$. Then we 
have that 
\begin{equation}
\label{eq.23}
\delta \bB = \kappa\star\bB \mapsto 
(-\delta)(\bB) = \kappa(-\star)(\bB) , 
\end{equation}
and no problem appears within the parity of $\bB$. The same holds 
for the equation $\nabla\times\vec{E} = \kappa\vec{E}$ 
which reads $d\bE = \kappa\star \bE$, and transforms as 
\begin{equation}
\label{eq.25}
d\bE = \kappa\star\bE \mapsto (-d)(-\bE) = 
\kappa(-\star)(-\bE) .
\end{equation}

In summary, when correctly formulated in terms of differential 
forms, that is, the electric field being represented by a 1-form 
and the magnetic field being represented by a 2-form, 
the vector equation $\nabla\times\vec{a} = \kappa\vec{a}$ does 
not show any problem related to violation of 
parity invariance. 

Moreover, since the calculus with differential forms is 
{\it intrinsic\/} \cite{Schultz}, it does {\it not\/} 
depend on our coordinate system choice. We remember, however, 
that the vector equations $\nabla\times\vec{E} = 
\kappa\vec{E}$ and $\nabla\times\vec{B} = \kappa\vec{B}$ 
emerged from a separation of variables which is expected to 
hold only in the rest frame. 

In conclusion, when correctly formulated, the vector 
equation $\nabla\times\vec{a} = \kappa\vec{a}$ does 
not deserve any of Salingaros' criticisms \cite{Saling}. 

Before we end we recall that being $\langle x^\mu \rangle $ 
($\mu = 0,1,2,3$) Lorentz coordinates of Minkowski spacetime, 
the Maxwell equations can be written as 
\begin{equation}
\label{eq.26}
d \bF = 0 , \quad \delta \bF = -\bJ , 
\end{equation}
where $\bF = (1/2)F_{\mu\nu}dx^\mu \w dx^\nu$ and 
$\bJ = J_\mu dx^\mu$, with 
\begin{equation}
\label{eq.27}
F_{\mu\nu} = \left( \begin{array}{cccc}
                     0 & E_1 & E_2 & E_3 \\
                     -E_1 & 0 & -B_3 & B_2 \\
                     -E_2 & B_3 & 0 & -B_1 \\
                     -E_3 & -B_2 & B_1 & 0 
                     \end{array} \right) , \quad 
J_\mu = \left( \rho , -j_1 , -j_2 , -j_3 \right) . 
\end{equation}
The force-free equation appears, e.g., in the tentative to 
construct purely electromagnetic particles (PEP), as done, 
for example, in \cite{Waite,Waite1}. Following Einstein\cite{Ein}, 
Poincar\'e\cite{Poin} and Ehrenfest\cite{Eh} a PEP must be 
free of self-force. Then the current vector field 
$\underline{J} = J^\mu \partial_\mu $ must satisfy 
\begin{equation}
\label{eq.28}
\underline{J} \JJ \bF = 0 , 
\end{equation}
or in vector notation, 
\begin{equation}
\label{eq.29}
\rho\vec{E} = 0 , \quad \vec{j}\cdot\vec{E} = 0 , 
\quad \vec{j}\times\vec{B} = 0 . 
\end{equation}
From eq.(\ref{eq.29}) Einstein concluded that the only 
possible solution of eq.(\ref{eq.26}) with the condition 
given by eq.(\ref{eq.28}) is that $\underline{J} = 0$. 
However, this conclusion only holds if we assume that 
$\underline{J}$ is time-like. If we assume that $\underline{J}$ 
may be space-like (as, for example, in London's theory of 
superconductivity) then there exists a reference frame where 
$\rho = 0$, and a possible solution of eq.(\ref{eq.28}) 
is 
\begin{equation}
\label{eq.30}
\rho = 0 , \quad \vec{E}\cdot\vec{B} = 0 , \quad 
\vec{j} = k  C \vec{B} , 
\end{equation}
where $k = \pm 1$ is called the chirality of the solution                   
and $C$ is a real constant. In \cite{Waite,Waite1} 
stationary solutions of eq.(\ref{eq.26}) with the 
condition (\ref{eq.28}) are exhibited with $\vec{E} = 0$. 
In this case we verify that 
\begin{equation}
\label{eq.31}
\nabla\times\vec{B} = k C \vec{B} .
\end{equation}
What is interesting to observe is that from the solutions 
of eq.(\ref{eq.31}) found in \cite{Waite,Waite1} we can obtain 
solutions of the free Maxwell equations. Indeed, it is enough 
to put $\vec{E}^\prime = \vec{B}\cos{\Omega t}$ and 
$\vec{B}^\prime = \vec{B}\sin{\Omega t}$, as discussed in the 
beginning. In \cite{Conf} we found also stationary solutions 
of Maxwell equations. Other solutions can be found with the 
methods described in \cite{Wal}.

\end{document}

%% file: climacro.tex
\newcommand{\JJ}{\mathbin{\raisebox{0.25ex}{$\footnotesize
                       \rm\vphantom{I}%
                       \_\hskip -0.25em\_%
                       \vrule width 0.6pt$}}}         

\newcommand{\w}{\wedge}

\newcommand{\bB}{{\bf B}}  \newcommand{\bb}{{\bf b}}

\newcommand{\bE}{{\bf E}}    
\newcommand{\bF}{{\bf F}}

\newcommand{\bJ}{{\bf J}}

  \newcommand{\bu}{{\bf u}}
  \newcommand{\bv}{{\bf v}}
  \newcommand{\bw}{{\bf w}}

\newcommand{\cl}{C \kern -0.1em \ell}     
            
\newcommand{\subB}[1]{!\raise-5pt\hbox{\mathsurround 0pt $#1$}}

\newcommand{\abs}[1]{\mathopen{\vert} #1\mathclose{\vert}}

\newcommand{\ut}[1]{{\setbox0=\hbox{$#1$}\mathsurround=0pt  
       \rlap{\raisebox{-0.8\dp0}{\raisebox{-0.8ex}
       {\kern -0.15ex\hbox{$\tiny\sim$}\kern 0.15ex}}}#1}}
\newcommand{\uti}[1]{{\setbox0=\hbox{$#1$}\mathsurround=0pt 
       \rlap{\raisebox{-0.8\dp0}{\raisebox{-0.8ex}          
       {\kern -0.3ex\hbox{$\tiny\sim$}\kern 0.3ex}}}#1}}

\newdimen\arrayruleHwidth                 
     \makeatletter
     \def\Hline{\noalign{\ifnum0=`}\fi\hrule \@height \arrayruleHwidth
         \futurelet \@tempa\@xhline}
     \makeatother

\newcommand{\BR}{{\rm\hskip 0.1pt
                       I\hskip -2.15pt R}}